\documentclass[referee]{raa}            

\usepackage{graphicx,times}             
\usepackage[T1]{fontenc}
\usepackage{natbib}
\usepackage{amssymb,amsmath}
\bibpunct{(}{)}{;}{a}{}{,}
\usepackage[pagebackref=true]{hyperref}

\begin{document}

  \title{Deuteration of water in protoplanetary discs during luminosity outbursts: model predictions for FU~Ori discs
}

   \volnopage{Vol.0 (20xx) No.0, 000--000}      
   \setcounter{page}{1}          
   \author{Anastasiia Topchieva 
      \inst{1,2}
   \and Tamara Molyarova
      \inst{3}
   \and Anton Vasyunin
      \inst{2}
   }

   \institute{Institute of Astronomy, Russian Academy of Sciences, 
             Pyatnitskaya St. 48, 
             Moscow, 119017, Russia; {\it ATopchieva@inasan.ru}\\
        \and
             Ural Federal University named after the first President of Russia B.N. Yeltsin, Mira Av. 19, Yekaterinburg, 620002, Russia\\
        \and
             Research Institute of Physics, Southern Federal University, Rostov-on-Don, Russia\\
\vs\no
   {\small Received 20xx month day; accepted 20xx month day}}

\abstract{ Luminosity outbursts of FU~Ori-type objects (FUors) allow us to observe in the gas the molecules that are typically present in the ice in protoplanetary discs. In particular, the fraction of deuterated water, which is usually is mostly frozen in the midplane of a protoplanetary disc, has been measured for the first time in the gas of the disc around a FUor V883~Ori.
We test the hypothesis that the observed high HDO/H$_{2}$O ratio in the V883~Ori protoplanetary disc can be explained by luminosity outbursts of different amplitude, including a series of two consecutive outbursts. Using the ANDES astrochemical code, we modelled the distributions of water and deuterated water abundances under the action of luminosity outbursts of different amplitudes (from 400 to 10\,000\,$L_{\odot}$) and at different stellar luminosities at the pre-outburst stage. We show that the best agreement with the observed HDO/H$_{2}$O profile is obtained for outburst amplitudes of 2\,000 and 10\,000\,$L_{\odot}$, while the observed bolometric luminosity of V883~Ori does not exceed 400\,$L_{\odot}$. We discuss possible reasons for this discrepancy, including the presence of past luminosity outbursts, the age of the star, and the influence of additional heating mechanisms in the midplane of the protoplanetary disc. We also consider how the high observed $\rm HDO/H_{2}O$ ratio may be related to the evolution of the chemical composition of the ice in the protoplanetary disc and the chemical processes activated under outburst conditions.
$\cdots\cdots$
}

   \authorrunning{A. Topchieva, T. Molyarova\& A. Vasyunin }            
   \titlerunning{Deuteration in Protoplanetary Discs}  

   \maketitle

\section{Introduction}           
\label{sect:intro}

The study of the water ice in protoplanetary discs is key to understanding the origin of water in planetary systems, including the Earth~\citep{2025PhyU...68..278K}. Water in protoplanetary discs plays a key role in the chemical evolution and formation of planets~\citep{2017A&A...608A..92D,2021A&A...650A.185M}. Accretion outbursts, such as those observed in FU~Ori-type stars (FUors), can significantly affect the distribution of water and its deuterated form (HDO). The chemistry of deuterated species differs from that of the main isotopologues. Being sensitive to values of medium temperature and density as well as radiation, the ratio between H$_2$O and HDO abundances must evolve during star formation from the interstellar cloud to the protoplanetary disc and the forming planetesimals and planets~\citep{Cleeves2014, 2016A&A...586A.127F}.

The modelling results show that during luminosity outbursts, the temperature in the protoplanetary disc increases dramatically, leading to the evaporation of water ice and a change in the HDO/H$_2$O ratio in the gas phase. This can be reflected in the observed spectra, providing information on the dynamics and chemistry of the protoplanetary disc \citep{Tobin2023, 2019MNRAS.485.1843W}. In particular, a high HDO/H$_2$O ratio may indicate the origin of water in the cold molecular phase before star formation, while its decrease is possibly due to thermal processing in the hot inner regions of the protoplanetary disc \citep{2013Icar..223..722J, Tobin2023}.

Modern interferometric observations allow us to study the spatial distribution of HDO and H$_2$O in the discs and envelopes of young protostars. In particular, for Class~0 and Class~I objects, such as V883~Ori and L1551~IRS5, the HDO/H$_2$O ratios obtained from observations reach $(2.1{-}2.3)\times10^{-3}$, which is significantly higher than the values typical of terrestrial oceans ($\sim1.6\times10^{-4}$) and comparable to the observed values in comets \citep{Andreu2023, Tobin2023}. These results support the hypothesis of ``inherited'' water that has passed from the molecular cloud into the protoplanetary disc without significant chemical processing.

At present, the evolution of the HDO/H$_2$O ratio under conditions of variable luminosity remains an open question. FUors, which V883~Ori and L1551~IRS5 are examples of, undergo accretion outbursts accompanied by a sharp increase in luminosity to hundreds or thousands of $L_{\odot}$.

Observations of the V883~Ori protoplanetary disc in the HDO emission line show that the water snow line is located at a distance of $\sim$80\,au from the star~\citep{Tobin2023}. Indirect methods give different but also high values for this distance: $\sim$40\,au from dust emission from \citep{2016Natur.535..258C} and $\sim$100\,au from HCO$^+$~\citep{2021A&A...646A...3L}. At the same time, under the standard luminosity of the star ($\lesssim10$\,$L_{\odot}$), the position of the snow line is expected to be no more than a few au away, and even when the luminosity increases to $\sim400$\,$L_{\odot}$, radiative heating is able to shift the water snow line to a maximum distance of $\sim20$\,au~\citep{2015A&A...582A..41H}. This discrepancy suggests that residual heating from previous outbursts or additional mechanisms of heat redistribution, such as non-radiative heating in the midplane of \citep{2024MNRAS.527.9655A}, need to be taken into account.

Although early models of \citep{Furuya2013, 2007prpl.conf..751B} show that vertical mixing and photochemistry play an important role in the redistribution of HDO/H$_2$O in the protoplanetary disc, they did not consider the influence of FUor luminosity outbursts, whose brief heating can also cause changes in the relative abundance of deuterated water \citep{2015MNRAS.446.3285O}. We consider a scenario in which it is luminosity outbursts, rather than slow mixing processes, that are the key factor in shaping the observed deuterated water distribution.

In the present work, we investigate the effect of accretion outbursts on the evolution of the HDO/H$_2$O ratio in protoplanetary discs. The distributions of water and its deuterated form at different luminosity levels ($400{-}10\,000$\,$L_{\odot}$) are modelled using the ANDES astrochemical code to reproduce conditions before, during and after the outburst. Special attention is paid to the following aspects:
\begin{itemize}
\item whether the outburst can explain the observed high HDO/H$_2$O;
\item how the radial distribution of HDO/H$_2$O changes due to the outburst;
\item how much the position of the water snow line and the HDO/H$_2$O distribution depend on the outburst amplitude and the presence of additional heating mechanisms;
\item and to what extent the observed HDO/H$_2$O ratio reflects the composition of the original ice, or has undergone reprocessing during chemical evolution.
\end{itemize}

Our goal is to try to explain the existing ALMA data for V883~Ori and to determine which outburst scenario is most consistent with the observed HDO/H$_2$O ratio. In addition, we aim to establish a link between conditions in protoplanetary discs around FUors and the chemical composition of ice in the early the Solar system, including data from comets. This will improve our understanding of the role of accretion outbursts in the evolution of molecular composition and help to refine scenarios for the transport of water to Earth-like planets.

The paper is organised as follows. Section~\ref{sec:model} describes the numerical model and calculation parameters. Section~\ref{sec:resalts} presents the HDO/H$_2$O radial profiles for different outburst scenarios. Section~\ref{sec:discussion} discusses possible reasons for the discrepancy between models and observations, as well as the possible evolution of HDO/H$_2$O from the protoplanetary disc formation to the present-day Solar system. In Section~\ref{sec:conclusion}, we formulate the main conclusions regarding the effect of outbursts on the chemistry of water ice in protoplanetary discs.

\section{Model}
\label{sec:model}

\subsection{Observed parameters of the V883~Ori system and modelling the protoplanetary disc structure}
\label{sec:dust_growth}

In our modelling, we rely on the observed characteristics of the V883~Ori system. It has been observed many times with ALMA and other instruments, which allowed us to obtain estimates of its main parameters \citep{2016Natur.535..258C,2018MNRAS.474.4347C}. The star has a mass of about 1.3 $M_{\odot}$, the protoplanetary disc radius of 67~to 370\,au, and the protoplanetary disc mass of 0.2~to 0.38\,$M_{\odot}$, according to data from  \citet{2021ApJS..256...30K}. According to these data, the protoplanetary disc is quite massive: 20 to 50\% of the mass of the central star, suggesting the possibility of gravitational instability.  However, spatially resolved observations of dust emission show no evidence of asymmetric substructures characteristic of advanced gravitational instability~\citep{2016Natur.535..258C}. Newer observations indicate a lower value of the protoplanetary disc mass of $0.02-0.09$\,$M_{\odot}$ \citep{2017A&A...605L...2S,Tobin2023}, which is in better agreement with the absence of substructures in the disc. In the model, the mass of the protoplanetary disc is assumed to be $0.05$\,$M_{\odot}$, and the characteristic radius is assumed to be 125\,au.

The bolometric luminosity of V883~Ori is about 400\,$L_{\odot}$ \citep{2001ApJS..134..115S}. With newer distance measurements, the current luminosity is estimated to be lower \citep[e.g., $200$\,$L_{\odot}$][]{2016ApJS..224....5F}. We will rely on a conservative estimate of the current luminosity of 400\,$L_{\odot}$. As indicated by \citet{1993ApJ...412L..63S}, a brighter nebula around this star was observed as early as in 1888 by \citep{1890AnHar..18....1P}, indicating that the outburst started back then \citep[see, for example,][]{2018ApJ...861..145C, Tobin2023}. We will assume that the duration of the outburst is $\sim130$\,years. The age of the star is estimated to be about 0.5\,Myr; it belongs to Class~I of young stellar objects because it retains an envelope mass comparable to the mass of the protoplanetary disc~\citep{2009ApJS..181..321E,2016Natur.535..258C}. 
Indirect observations of the snow line in this protoplanetary disc at a large distance from the star suggest that the luminosity may have been higher in the past~\citep{2021A&A...646A...3L}. This is also indicated by historical observations of a brighter nebula in the vicinity of this star~\citep{2018ApJ...861..145C}. We therefore consider different scenarios for the luminosity change described below in~\ref{sec:outburst_model}.

We use the ANDES astrochemical code \citep{2013ApJ...766....8A} to study the chemical evolution of the protoplanetary disc under the influence of the FUor outburst. The structure of the disc is constructed following the model presented in \citet{2017ApJ...849..130M}. The protoplanetary disc is assumed to be axisymmetric, and the surface density distribution is described by a power law. The vertical structure is calculated from the hydrostatic equilibrium conditions.

The temperature of the protoplanetary disc, both in the radial and vertical directions, is determined using a combined method: for the atmosphere, the UV radiative transfer calculation is used, while the temperature in the midplane is determined parametrically based on the total source luminosity (stellar and accretion) \citep{2014ApJ...788...59W, 2016PhRvL.117y1101I, 2017ApJ...849..130M}. The density and temperature distributions are calculated iteratively until mutual agreement is achieved. We assume that dust and gas temperatures are equal, which is a reasonable approximation in the collisionally dominated disc midplane \citep{2004ApJ...615..991K,2009A&A...501..383W}. The optical properties of the dust are adopted from \citet{1993ApJ...402..441L}, with the dust mass assumed to be 1\% of the gas mass. We assume a power-law size distribution of dust grains with a $-3.5$ exponent and minimum and maximum dust grain sizes $a_{\rm min}=5\cdot 10^{-7}$\,cm and $a_{\rm max}=2.5\cdot 10^{-3}$\,cm, which corresponds to slightly grown dust compared to the standard values for the interstellar medium. A more detailed description of the dust model is given in Section~2.2 of~\citet{2018ApJ...866...46M}.

In order to model a Class~I object, it is necessary to take into account the surrounding envelope in addition to the protoplanetary disc itself. We consider a model based on the work of \citet{2003ApJ...591.1049W}, which describes the density distribution in a rotating accreting envelope. A more detailed implementation of the envelope in the ANDES model and its influence on the chemistry of FUors is described in \citet{2019INASR...4...45M,2024MNRAS.527.7652Z}. The main parameters of the envelope in the model are the outer radius (1000\,au), the accretion rate through the envelope $10^{-5}$\,$M_{\odot}$\,year$_{-1}$, and the centrifugal radius equal to the characteristic radius of the protoplanetary disc 125\,au. With the parameters given in the model, the total mass of the envelope is $0.026$\,$M_{\odot}$.

\subsection{Chemical model}
\label{sec:chemical_model}

To describe the chemical evolution of deuterated water, we use a chemical reaction network that includes deuterium fractionation \citep{2011ApJS..196...25S, 2013ApJS..207...27A}. It includes 1247 chemical components and 38347 reactions, including gas-phase and surface two-body reactions, adsorption and reactive desorption, photoreactions,  ionisation and dissociation by X-rays, cosmic rays and radioactive nuclides \citep[see, for example][]{2009A&A...501..383W, 2013ApJ...766....8A}. The considered set of species includes mono-, di-, and trideuterated molecules and ions, including those on dust surface. The model also includes ortho- and para-isomers of molecular hydrogen, which is necessary to correctly describe the chemistry of deuterated species.

The model is optimised and tested for the cold conditions of the interstellar medium and reproduces well the chemistry of deuterated species at temperatures below 150\,K~\citep{2013ApJS..207...27A}. We suggest that the water vapour composition observed in the protoplanetary disc around V883~Ori is mainly determined by the composition of the ice evaporated by the luminosity outburst. Gas-phase two-body reactions for higher temperatures are available in the reaction network and are accounted for temperatures up to 600\,K. Also included in the grid are photoreactions that can influence the dissociation of water in the molecular layer. However, it should be noted that the results of calculations in the inner protoplanetary disc and at temperatures above 600\,K are less reliable. This feature of the model nevertheless does not prevent comparison with the observed composition of the protoplanetary disc: spatially resolved data for the radial concentrations of water isotopologues are obtained for distances $>40$\,au, and the temperature in the protoplanetary disc midplane at this distance does not exceed 250\,K even at a luminosity of 10\,000\,$L_{\odot}$.

As an initial chemical composition, we use the composition of ices observed in protostellar cores~\citet{2016A&A...595A..83E, 2011ApJ...740..109O}, where molecular ices of the main volatile species are present. At the initial time instant, deuterium is present only in the gas phase as part of the HD molecule, its initial abundance being $1.55 \cdot 10^{-5}$ with respect to hydrogen nuclei. This value corresponds to the observed deuterium abundance in the interstellar medium~\citep{1994ARA&A..32..191W}.

\subsection{Luminosity outburst model}
\label{sec:outburst_model}

We evolve disc physical structure due to the effect of an outburst following the approach from \citet{2018ApJ...866...46M}. The total luminosity of the central source consists of the intrinsic stellar luminosity and the accretion luminosity, which changes throughout the outburst. Changes in the total luminosity are reflected in the vertical structure of the disc, which is recalculated at each new luminosity value. In this work, we do not consider the hydrodynamical response of the protoplanetary disc to the outburst and assume that the structure adapts instantaneously to the change in luminosity. As the accretion luminosity increases, the protoplanetary disc heats up, thickens, and becomes more inflated. It should be noted that the models are quasi-stationary and not all regions of the protoplanetary disc manage to reach an equilibrium state within a typical outburst time (on the order of 100 years), see, e.g. equations 24-30 in \citet{1997ApJ...490..368C} or Figure~3 in \citet{2014ARep...58..522V}.
The effects of thermal inertia and gas dynamics are not accounted for in the current model; inclusion of these processes may further shift the snow line in the late stages of the outburst~\citep[see][]{2014ARep...58..522V}.

\begin{figure}
\includegraphics[width =0.49\columnwidth]{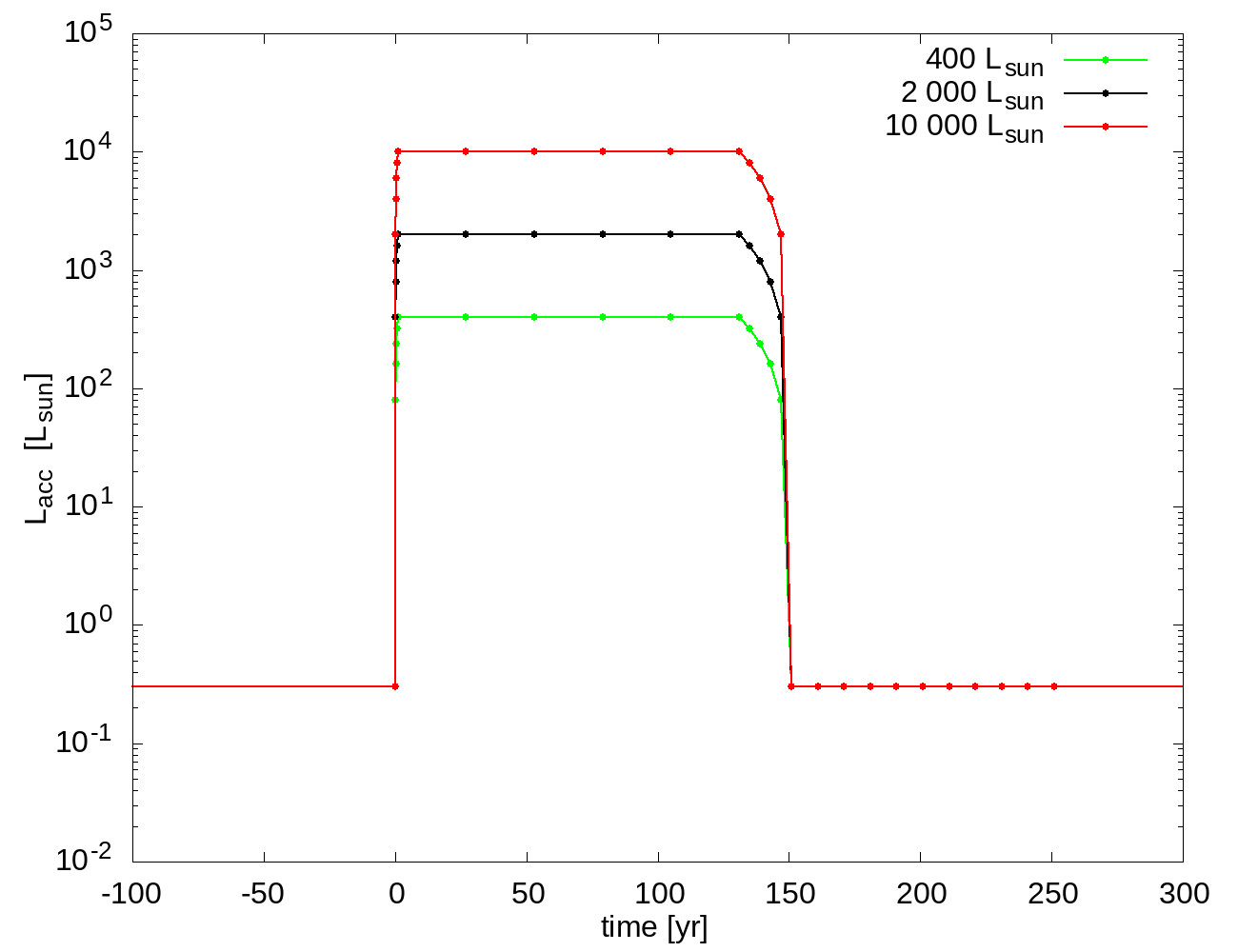}
\includegraphics[width =0.49\columnwidth]{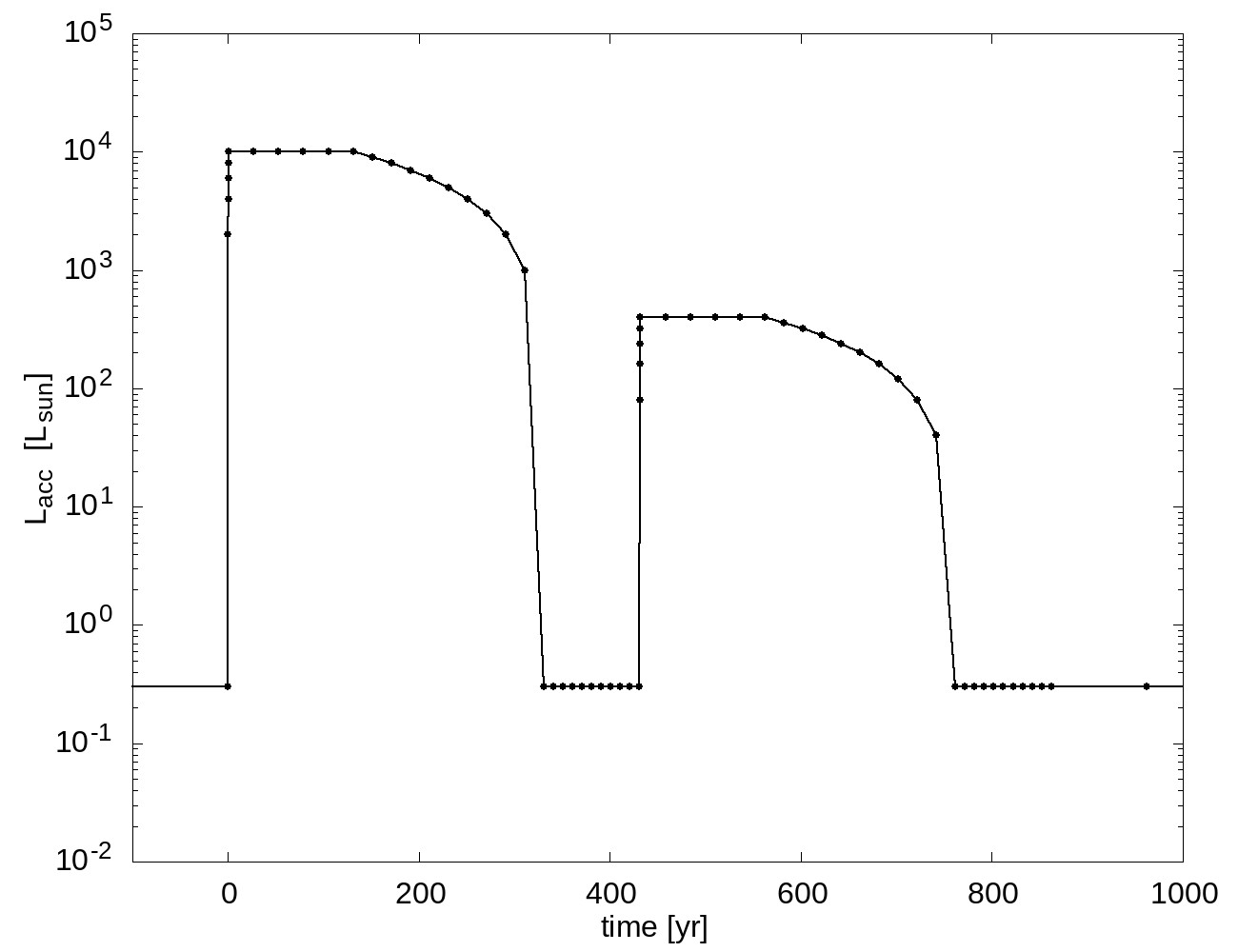}
\caption{Evolution of the total luminosity of the accretion source in different models. Three models with different outburst amplitudes: 400, 2\,000, and 10\,000 $L_{\odot}$ are shown on the left. On the right is the model with two consecutive outbursts with amplitudes of 10\,000 and 400 $L_{\odot}$. The time is specified relative to the outburst onset.}
\label{ris:grid}
\end{figure}

We assume that the outburst lasts 130~years, based on the observed duration of the V883~Ori outburst~\citep{2018ApJ...861..145C}. Before the onset of the outburst, the accretion luminosity remains unchanged at $0.3$\,$L_{\odot}$, which is a typical value for young T~Tauri-type stars. With the onset of the outburst, the luminosity rises to its maximum amplitude during one year. We consider outbursts with amplitudes of 400, 2\,000, and 10\,000\,$L_{\odot}$, as well as the case where two consecutive outbursts occur. The luminosity is assumed to have a maximum value for 130~years, after which it declines to a pre-outburst level for 20~years. Model profiles of the evolution of the accretion luminosity for several models with different peak luminosities are shown in Fig.~\ref{ris:grid}.

In the pre-outburst phase, the star is the main contributor to the luminosity. Ages of young stars are not precisely measured, but luminosity of the star during the pre-outburst period can influence the composition of the ice exposed by the outburst. The age of V883~Ori is considered to be $\approx0.5$\,Myr\citep[see][]{2016Natur.535..258C}, but this estimate is based mainly on the presence of a surrounding envelope and thus the object belonging to Class~I. Estimates of the duration of the evolutionary stages of young stars show that they may be in the class~0 stage for $\sim0.1$\,Myr and in the Class~I stage for $\sim0.5$\,Myr~\citep{2009ApJS..181..321E}. The luminosity of a star $L_{\star}$ with the adopted mass of $1.3$\,$M_{\odot}$ is $\sim17$\,$L_{\odot}$ at an age of $0.1$\,Myr and $\sim5$\,$L_{\odot}$ at an age of $0.5$\,Myr, according to the evolutionary tracks from the \citet{2008ASPC..387..189Y} model. We consider models with these two values of stellar luminosity.

However, we do not consider different values of the outburst duration. As shown in \citet{2019MNRAS.485.1843W}, for outbursts longer than 100\,years, the outburst duration does not significantly affect the chemical composition of the disc.

\begin{table}
\begin{center}
\caption[]{ Modelling parameters for accretion outbursts.}\label{Tab1}

 \begin{tabular}{clc}
  \hline\noalign{\smallskip}
No &  \textbf{$L_{\mathrm{acc}}$, $L_{\odot}$} &  \textbf{Features}  \\
  \hline\noalign{\smallskip}
1 & 400  & Single outburst, observed amplitude  \\
2 & 2\,000  & Single outburst, increased amplitude \\
3 & 10\,000 & Single outburst, maximum amplitude \\
4 & 10\,000 + 400 & Two consecutive outbursts (100\,yr apart): first 10\,000~$L_{\odot}$, then 400~$L_{\odot}$ \\
  \noalign{\smallskip}\hline
\end{tabular}
\label{tab:burst_parameters}
\end{center}
\end{table}

Table~\ref{tab:burst_parameters} presents the parameters of the numerical models of accretion outbursts used in this work. 

For each model, the time interval before its onset is $\Delta t=0.5$\,Myr, and two values of the stellar luminosity are considered.
We consider both single outbursts with different amplitudes and time intervals before the onset of the outburst, as well as a scenario with two consecutive outbursts.

\section{Results}
\label{sec:resalts}
\subsection{Water snow line position}
\label{sec:water_snowline}

The temperature in the disc plane $T_{\rm mp}$ depends on the luminosity of the central source and increases during the accretion burst. In the model, it is calculated from the temperature and size of the star and the accretion region:
\begin{equation}
\label{eq:temp1}
    T_{\rm mp}^4(R) = \frac{\alpha}{2}\left( T_{\star}^4 \left(\frac{R_{\star}}{R}\right)^2 + T_{\rm acc}^4 \left(\frac{R_{\rm acc}}{R}\right)^2  \right)
\end{equation}

Here $\alpha=0.05$ is the angle between the surface of the protoplanetary disc and the light beam from the central source. This temperature profile describes an optically thick protoplanetary disc with an inner hole illuminated by a central source \citep{1997ApJ...490..368C}, \citep{2001ApJ...560..957D}.

The stellar temperature $T_{\star}$ and the stellar radius $R_{\star}$ in the model are set according to the evolutionary tracks from \citet{2008ASPC..387..189Y} and correspond to the stellar luminosities of $\approx5$ and $\approx17$\,$L_{\odot}$ for the two considered stellar ages. The temperature of the accretion region is assumed to be $T_{\rm acc} = 10\,000$\,K, and the radius is set based on a given value of accretion luminosity $L_{\rm acc} = 4\pi R_{\rm acc}^2\sigma_{\rm SB} T_{\rm acc}^4$, which varies with time (see Section~\ref{sec:outburst_model}). If we assume that the accretion luminosity is much larger than the luminosity of the star, which is true for the considered outbursts of luminosities $>100$\,$L_{\odot}$, we obtain $T_{\rm mp}(R)\sim L_{\rm acc}^{1/4} R^{-1/2}$ during the outburst.  We further discuss an alternative temperature profile in Section~\ref{sec:temperature2}.

\begin{figure} 
\includegraphics[width =0.8\columnwidth]{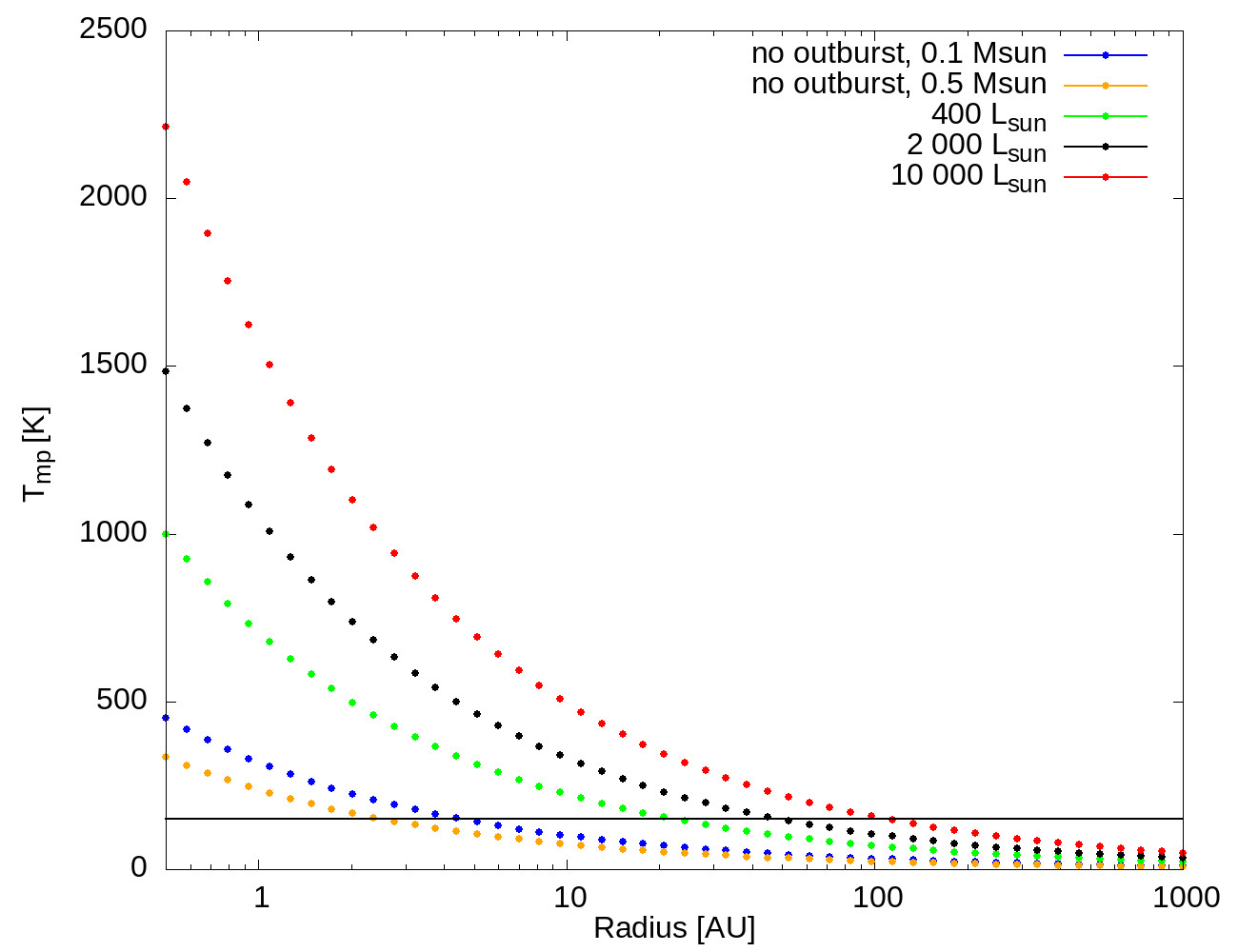}
\caption{Temperature in the protoplanetary disc midplane with no outburst for stellar ages of $0.1$ and $0.5$\,$M_{\odot}$ and at different values of accretion luminosity. The horizontal black line indicates the sublimation temperature of water ice ($\approx150$\,K). The intersections with it of the dashed temperature profiles correspond to the approximate position of the water snow line at the given luminosities.}
\label{ris:2TTT}
\end{figure}

During a luminosity outburst, the water snow line shifts further away from the star. Figure~\ref{ris:2TTT} shows the radial temperature distribution in the midplane under different conditions: in the absence of an outburst (before its onset) for stellar ages of $0.1$ and $0.5$\,$M_{\odot}$ and during an outburst with different luminosities. The characteristic sublimation temperature of water ice in protoplanetary discs is $\approx150$\,K, although this value depends on local environmental conditions, in particular on the gas density \citep{2015A&A...582A..41H}. The position of the water snow line in the midplane can be estimated as the distance at which the temperature is equal to this value. Above the midplane, temperatures are typically higher, and the snow line moves further away from the star and becomes a snow surface. However, observations of optically thin water emission allow us to determine the position of the snow line in the disc midplane \citep{Tobin2023}, so we focus on this temperature.

As can be seen from Figure~\ref{ris:2TTT}, at the quiescent stage the water snow line is at a distance of $\sim2-5$\,au, and it shifts to $\sim 20-100$\,au during an outburst. At the observed luminosity of V883~Ori $<400$\,$L_{\odot}$, the water snow line could be at a distance of $\sim 20$\,au (green dashed line in the figure~\ref{ris:2TTT}) if its position is determined only by radiative heating from the luminosity outburst. The distances $40-100$\,au obtained from observations~\citep{2016Natur.535..258C, 2021A&A...646A...3L, Tobin2023} should only be expected in this case for a much larger luminosity outburst, such as the $2\,000$ or $10\,000$\,$L_{\odot}$ shown here.

\begin{figure} 
\includegraphics[width =0.49\columnwidth]{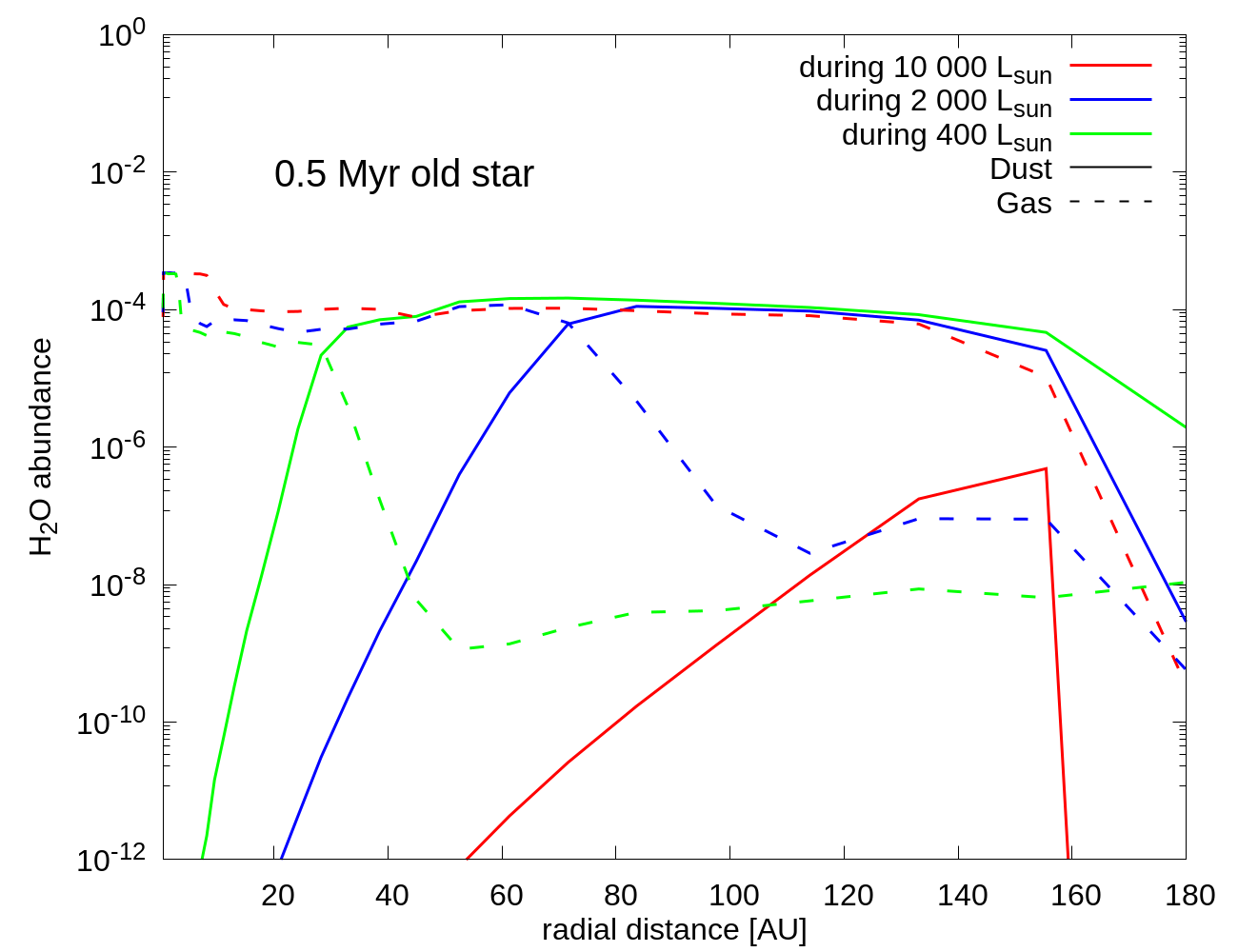}
\includegraphics[width =0.49\columnwidth]{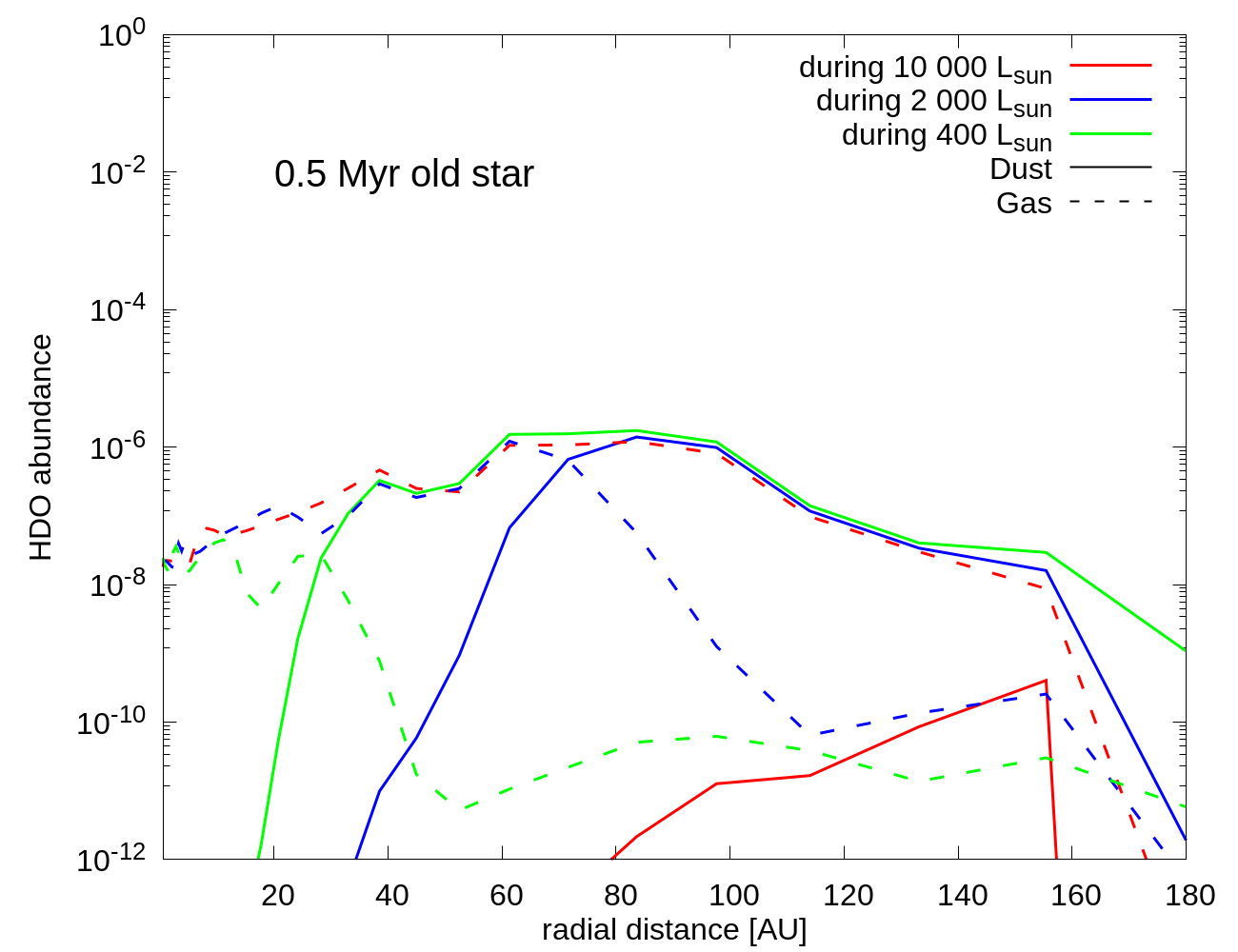}
\includegraphics[width =0.49\columnwidth]{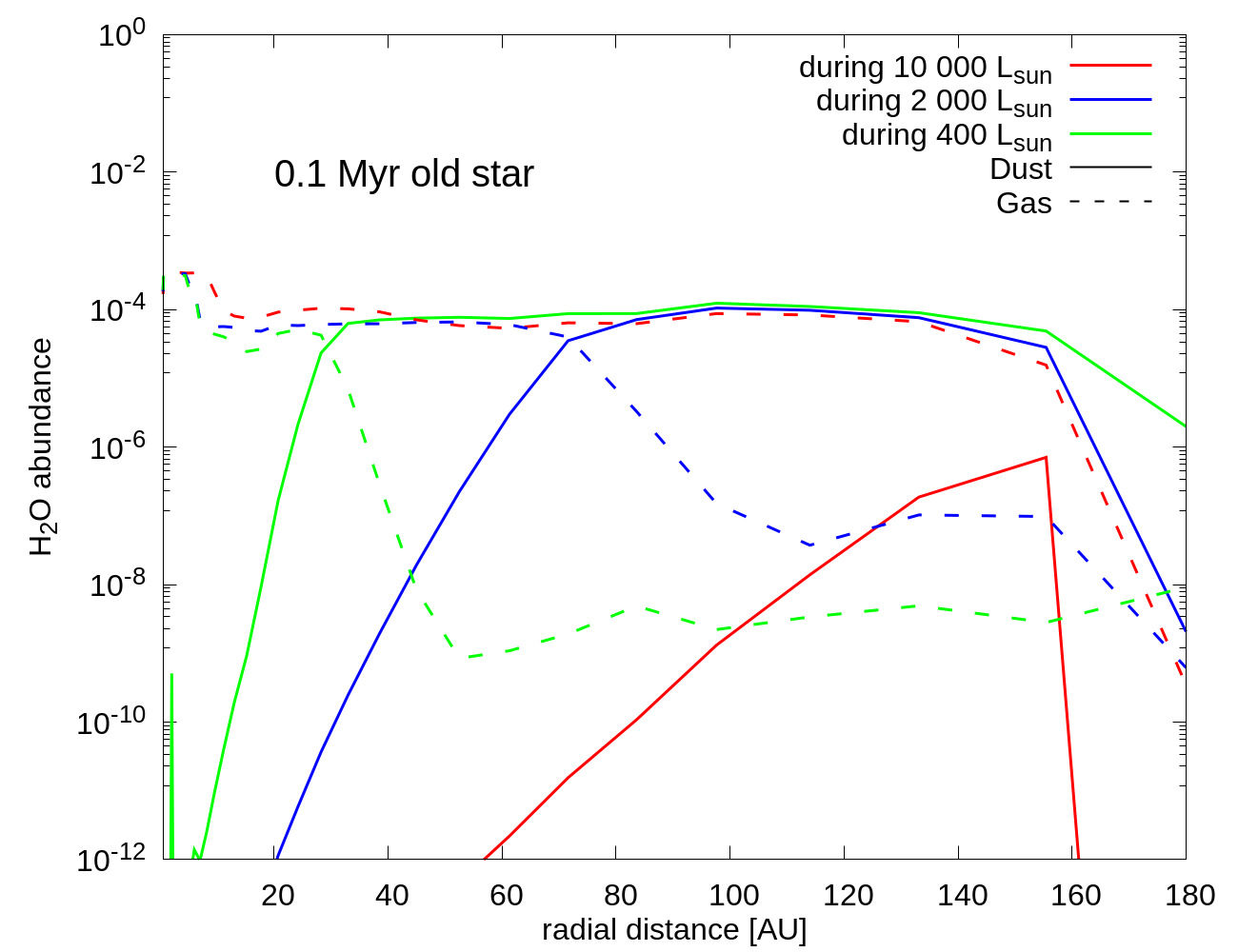}
\includegraphics[width =0.49\columnwidth]{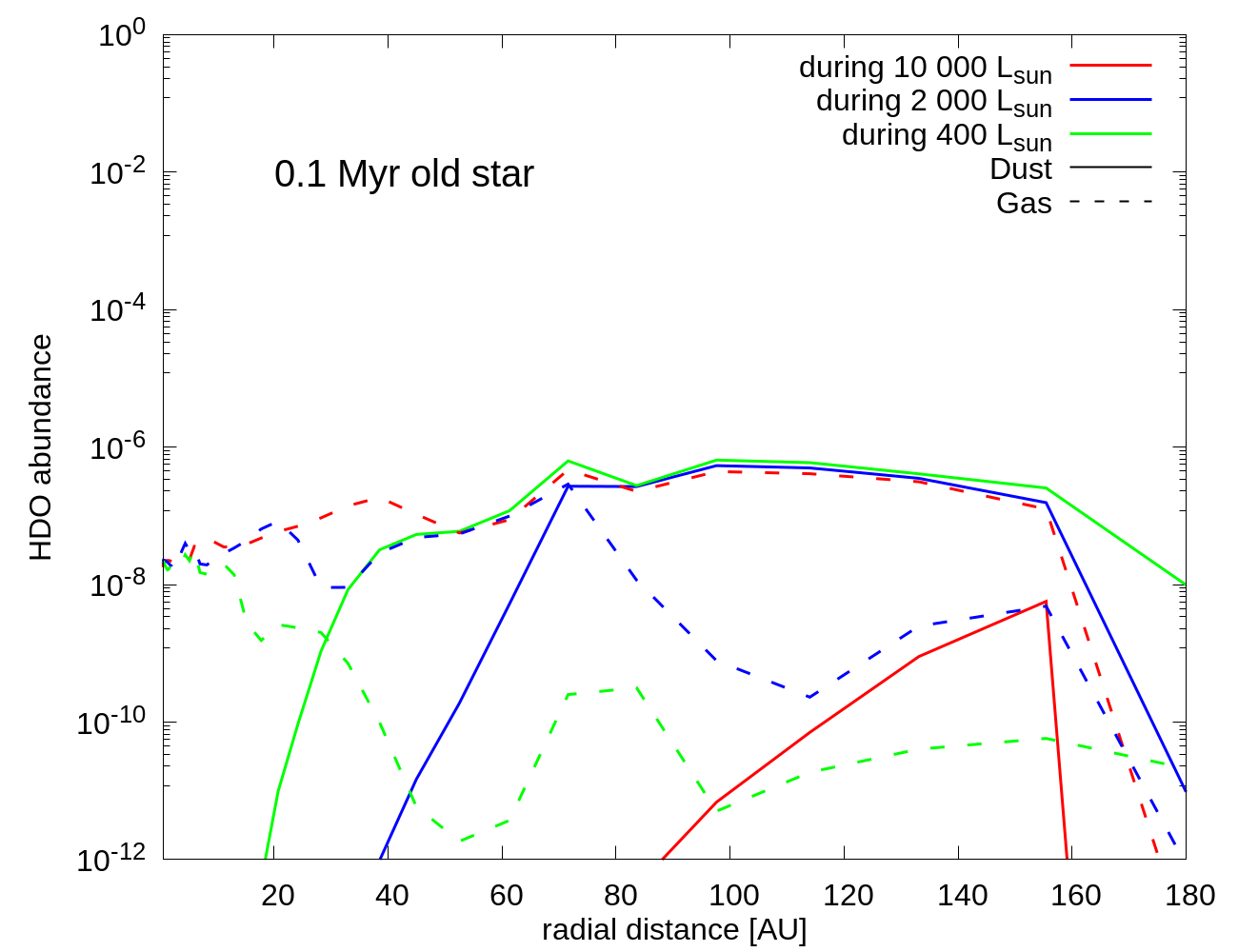}
\caption{Relative water abundance in the gas and on dust grain surface (in the ice) at different accretion luminosities (400, 2\,000, and 10\,000~$L_{\odot}$) for stellar ages of $0.1$ and $0.5$\,Myr. The red, green, and blue colours denote outbursts with luminosities of 10\,000, 2\,000, and 400~$L_{\odot}$, respectively. The intersection of ice and gas components defines the snow line. At 10\,000~$L_{\odot}$ (red), the ice abundance is low and the H$_2$O and HDO snow lines are absent. For fainter outbursts, the snow lines are at $20-30$\,au (400~$L_{\odot}$, blue) and $\sim70$\,au (2\,000~$L_{\odot}$, green).}
\label{ris:aqw}
\end{figure}

Figure~\ref{ris:aqw} shows the distributions of the relative abundances of H$_2$O and HDO in the gas and solid phases as a function of the distance to the star for models with three different outburst amplitudes (400, 2\,000, and 10\,000~$L_{\odot}$) and two stellar ages ($0.1$ and $0.5$\,Myr). The dotted lines correspond to the abundance in the gas phase, the solid lines shows the ice on dust grain surface. During a 10\,000~$L_{\odot}$ outburst (red lines), the temperature is so high that almost all water is converted to the gas phase and no snow line forms. For outbursts of 2\,000 and 400~$L_{\odot}$, characteristic intersections of the gas and solid components corresponding to the snow line positions are observed: about $20-30$\,au for 400\,$L_{\odot}$ and $60-70$\,au for 2\,000\,$L_{\odot}$.

In addition to thermal desorption, the adopted astrochemical model also includes photodesorption, which leads to the evaporation of water in the outer regions of the protoplanetary disc  more transparent for UV radiation. It reduces the ice abundance at distances $\gtrsim120$\,au. In addition, the model is two-dimensional $(R,z)$, that is, it considers the structure of the protoplanetary disc in the vertical direction, and the radial profiles shown in Figure~\ref{ris:aqw} are obtained by integration in the vertical direction. The freeze-out temperature of water in the upper layers of the protoplanetary disc is reached at greater distances from the star than in the midplane. When considering the full vertical span of the disc, it results in a shift of the snow line further away from the star and a smoother change in the gas and dust abundances in the vicinity of the snow line. In addition, the chemical kinetics model takes into account that the extinction rate depends on the gas density, and the freeze-out temperature can differ from the standard value of $150$\,K. The combination of these factors causes the snow line positions in the astrochemical model (Figure~\ref{ris:aqw}) to differ slightly from the estimate based on the freez-out temperature (Figure~\ref{ris:2TTT}).

The observed position of the snow line in the disc midplane at a distance of $40-100$\,au corresponds to a luminosity of the order of $10^3-10^4$\,$L_{\odot}$ and cannot be obtained at the current luminosity of $400$\,$L_{\odot}$ in the absence of heating mechanisms other than radiation heating and in a quasi-stationary disc. Perhaps, in the past, the snow line was shifted to these distances by a brighter outburst and has not since had time to come into agreement with the current luminosity due to the finite time of establishing the thermal balance. Different physical and chemical processes in the protoplanetary disc occur on different timescales, which may also explain the inconsistency of the snow line positions determined by different methods (from dust properties in \citet{2016Natur.535..258C}, from the emission of the chemical tracer HCO$^+$ in \citet{2021A&A...646A...3L}, from the emission of water isotopologues in \citet{Tobin2023}). In the following Section, we use astrochemical modelling to test whether the observed HDO/H$_2$O ratio is consistent with different outburst scenarios and whether, like the position of the water snow line, it indicates a brighter outburst.

\subsection{Comparing models with different luminosities}
\label{sec:results}

The model allows us to calculate the two-dimensional $(R,z)$ distribution of molecules in the protoplanetary disc evolving in time. We are interested in the ratio between the gas-phase abundances of the water isotopologues HDO and H$_2$O, namely, a comparison of this ratio with the observational data from \citet{Tobin2023}. Since the observations are spatially resolved only in the radial direction, we integrate the HDO and H$_2$O abundances over $z$, and consider the radial profiles of the HDO/H$_2$O ratio in comparison with the observations.

\begin{figure} 
\includegraphics[width =0.49\columnwidth]{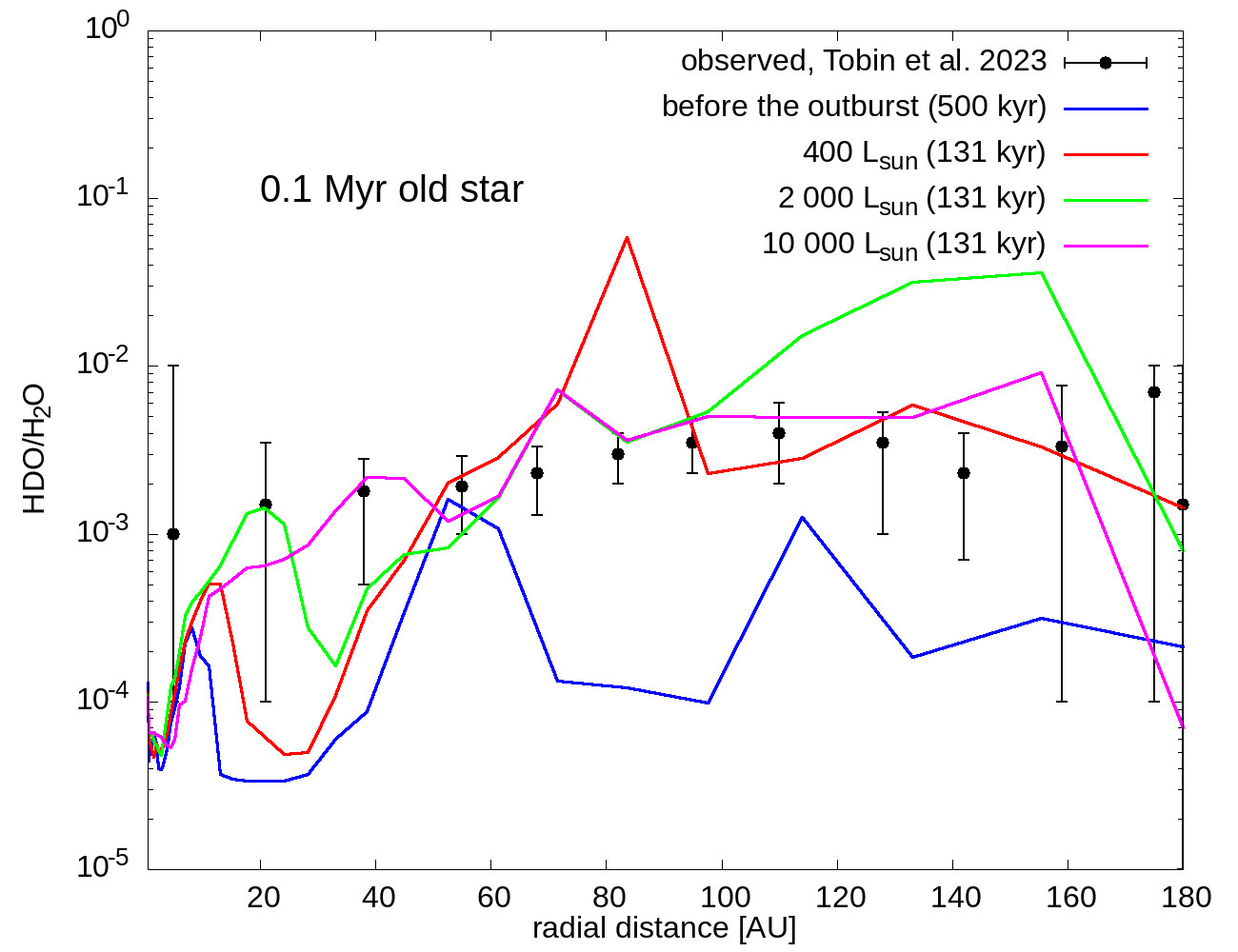}
\includegraphics[width =0.49\columnwidth]{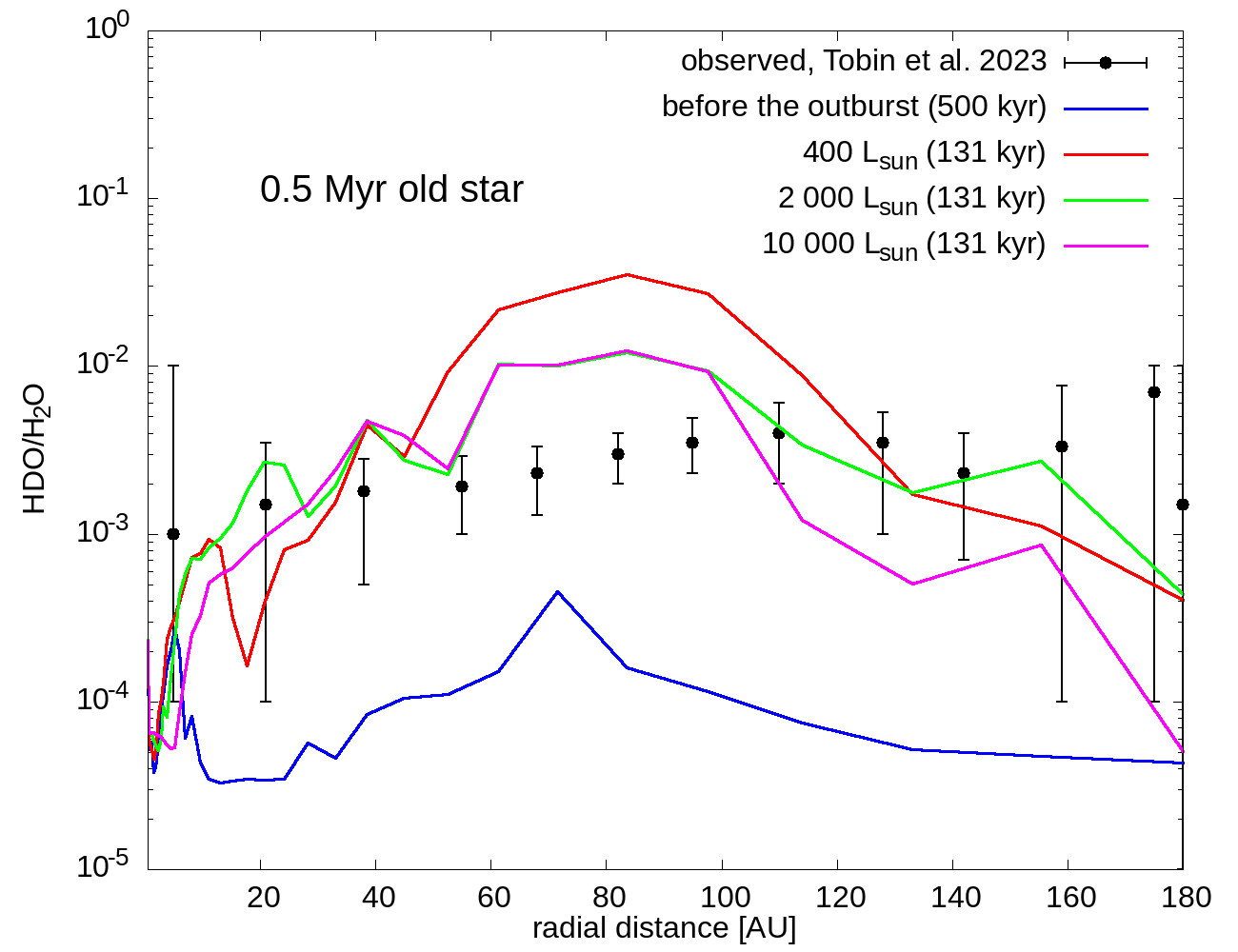}
\caption{Radial profiles of the HDO/H$_2$O ratio in the gas for different outburst scenarios. Black markers with error bars show observational data for the V883~Ori protoplanetary disc from \citet{Tobin2023}, with values inside 40\,au considered less reliable due to the contribution of the optically thick dust continuum. The blue line corresponds to the time before the outburst ($t=500$\,kyr). The red, green, and magenta lines show the simulation results 131~years after the outburst at different maximum luminosities: 400~$L_{\odot}$,  2\,000~$L_{\odot}$, and 10\,000~$L_{\odot}$, respectively.}
\label{ris:comb}
\end{figure}

Figure~\ref{ris:comb} compares the radial profiles of the HDO/H$_2$O ratio before and during the outburst in models with different peak luminosities and for two stellar ages ($0.1$~and $0.5$\,$M_{\odot}$). Prior to the outburst (500\,kyr time point, blue line in the figure~\ref{ris:comb}), the model predicts relatively low HDO/H$_2$O values across the disc. This ratio is determined mainly by molecules in the upper layers of the disc, where water (both H$_2$O and HDO) is in the gas phase due to photodesorption and higher temperatures. At the same time, most of water is in the disc midplane in form of ice mantles on the dust surface and does not contribute to the integrated gas-phase HDO/H$_2$O. During an outburst, this water ice evaporates; the region of the protoplanetary disc where this occurs depends on the brightness of the outburst. However, at all values of the maximum luminosity, water ice in both the upper layers of the protoplanetary disc and in the midplane will transition to the gas phase, consequently leading to a change in HDO/H$_2$O at all distances from the star. 
The water ice in the disc midplane evaporates and completely turns into the gas inside the water snow line (see Figure~\ref{ris:aqw}). At distances $\gtrsim40$\,au, the abundance of both H$_2$O and HDO in the gas phase at high luminosities almost completely coincides with the abundance of the corresponding ice at lower luminosities. This means that the gas-phase HDO/H$_2$O ratio within the snow line during the outburst is determined mainly by the ice composition before the outburst.

At 131\,yr, when the model luminosity is still at its maximum value (see Figure~\ref{ris:grid}), the HDO/H$_2$O ratio is elevated compared to the pre-outburst. This is particularly pronounced in the high peak luminosity models of 2\,000$L_{\odot}$ and 10\,000\,$L_{\odot}$. These models show better agreement with the observational data in the $60-100$\,au. The model with an outburst amplitude of 400\,$L_{\odot}$ shows a lower HDO/H$_2$O ratio in this region, except for a peak occurring at $\approx80$\,au for $0.1$\,Myr old star. The model with 400\,$L_{\odot}$ also has a higher hump in the region of $40-130$\,au than models with $2\,000$\,$L_{\odot}$ and $10\,000$\,$L_{\odot}$ outburst, for stellar ages of $0.5$\,Myr. This also supports the hypothesis of a brighter luminosity outburst in V883~Ori.

In the inner disc, at $r<40$ au, models with bright outbursts also demonstrate better agreement with observational data. However, as \citet{Tobin2023} points out, at these distances, dust emission is optically thick. This can lead to significant distortions in observed column densities and, consequently, to a more inaccurate observational HDO/H$_2$O ratio. When comparing models with observations, only distances greater than $40$\,au should be considered.

To interpret the local increase in the HDO/H$_2$O ratio during outbursts of different brightness, we analysed the chemical reactions contributing to the change in the HDO abundance in the gas phase. In the bulk of the disc, it is driven by ice evaporation and reflects the available HDO/H$_2$O in the ice phase in the protoplanetary disc prior to the outburst. However, there are also gas-phase reactions with different efficiencies for HDO and H$_2$O that contribute to the isotopologue ratio profile. In particular, in all models, there is a peak in the inner disc (at $10-40$\,au) that corresponds to the efficient formation of HDO in the molecular layer at $z/R\approx0.2$ (reaction $\mathrm{OH} + \mathrm{HD} \rightarrow \mathrm{HDO} + \mathrm{H}$). This peak shifts with increasing accretion luminosity due to an increase in the OH abundance due to an increase in the temperature and radiation field.

For a $0.5$\,Myr old star, which has lower luminosity before the outburst, there is an additional increase in the HDO abundance in both the gas and ice phases at a distance of $\sim60-100$\,au (see Figure~\ref{ris:aqw}). This is reflected as an increase in the HDO/H$_2$O ratio in the profiles in Figure~\ref{ris:comb}. In this region, the formation of water on dust grain surface is efficient, through  the reaction $\mathrm{GOH} + \mathrm{GH} \rightarrow \mathrm{GH_2O}$ and its deuterated analogues. The efficiency of deuterated water formation on dust is sensitive to the temperature and density \citep{2013A&A...550A.127T}, leading to differences in the HDO ice abundance in models with different stellar luminosity. 

Thus, the pre-outburst luminosity also affects the HDO/H$_2$O ratio in the ice phase, which is reflected in the gas-phase HDO/H$_2$O during the outburst due to water ice evaporation. In this paper we consider two different values of stellar luminosity, but in reality an evolving protostar gradually changes luminosity, which should lead to a change in thermal structure even in the absence of luminosity outbursts.

For the leas bright bright outburst (400\,$L_{\odot}$), the HDO/H$_2$O radial profile also shows a pronounced maximum at $\sim80$\,au, similar to the one that in the pre-outburst phase is located at $50-60$\,au. At more powerful outbursts (2\,000\,$L_{\odot}$), the maximum is less pronounced and shifts to the outer regions of the protoplanetary disc ($100-180$\,au), while at extreme luminosities (10\,000\,$L_{\odot}$) the HDO/H$_2$O enhancement has a more homogeneous structure. The reaction of HDO formation in the gas phase in the protoplanetary disc plane is responsible for the HDO/H$_2$O enhancement in these regions, which has the most contribution for the 400\,$L_{\odot}$ outburst of a 0.5\,Myr old star. HDO is formed from deuterated formaldehyde, HDCO, whose abundance is elevated relative to H$_2$CO. In turn, additional HDCO comes into the gas phase from the dust surface when the outburst vaporises the icy HDCO present in these regions. This reaction has lower contribution in the models with brighter outbursts.

We show that high HDO/H$_2$O values in FU~Ori-type discs can persist even after multiple outbursts, which is consistent with the idea of transfer of inherited ice from the molecular cloud. This behaviour may explain the observed elevated deuterium abundance in Hale-Bopp-type comets \citep[$3 \cdot 10^{-4}$][]{1998Sci...279..842M}, which confirms the connection between the early chemistry of protoplanetary discs and the composition of Solar system bodies. We obtain similar to these cometary values in the inner disc, and the outer discs have even higher values. 

\subsection{Outburst shape and the effect of past changes in luminosity}
\label{sec:double_outburst}

Although the current luminosity of V883~Ori is $\sim$400\,$L_{\odot}$, the luminosity required to reproduce the observed HDO/H$2$O profiles on the order of $2\,000-10\,000$\,$L{\odot}$ may have been achieved in a previous more powerful outburst. In particular, a bright outburst may have occurred in the system in the past, leaving a trace in the form of snow line positions and a fraction of deuterated water that persist even at lower current luminosities. We consider a model in which two consecutive outbursts occur with amplitudes of $10\,000$ and $400$\,$L_{\odot}$ and compare HDO/H$_2$O during these different outbursts. We also consider a later time point at a fading stage of the outburst in the model with $L_{\rm acc}=2\,000$\,$L_{\odot}$, when the accretion luminosity is also equal to $400$\,$L_{\odot}$. These scenarios represent possible options for the evolution of the luminosity. Since V883~Ori was not observed before the \citep{2018ApJ...861..145C} outburst, the detailed history of the luminosity evolution is unknown, and the proposed scenarios could have occurred.

\begin{figure} 
\includegraphics[width =0.49\columnwidth]{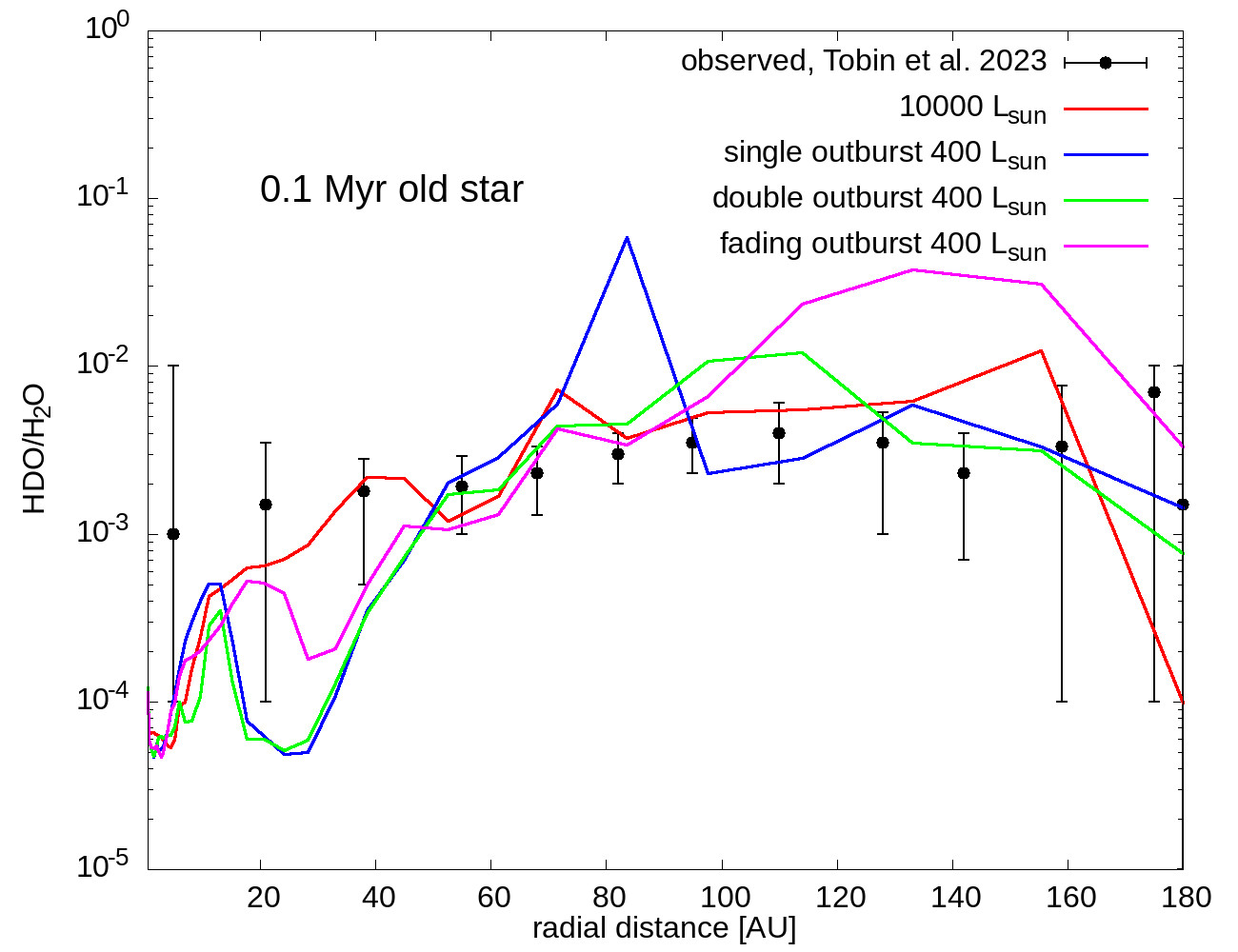}
\includegraphics[width =0.49\columnwidth]{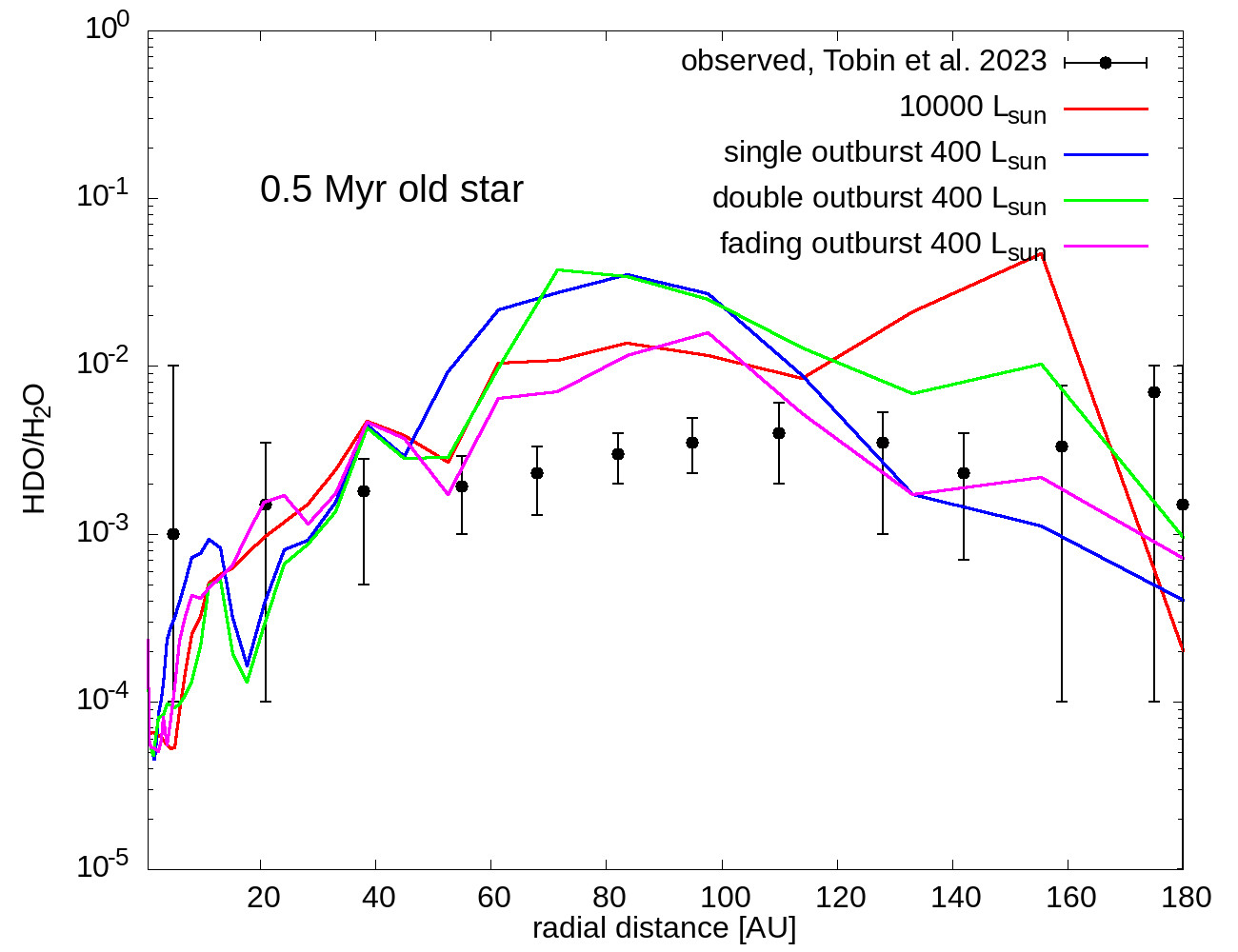}
\caption{Comparison of model HDO/H$_2$O radial profiles for stellar ages of $0.1$ and $0.5$ $M_{\odot}$ with observed data for V883~Ori \citep[][black markers with error bars]{Tobin2023}. The shown scenarios are: single 10\,000\,$L_{\odot}$ outburst (blue line), single 400\,$L_{\odot}$ outburst (green line), the second outburst in the two $10\,000+400$\,$L_{\odot}$ outbursts scenario (red line), a fading 2\,000\,$L_{\odot}$ outburst at the moment when the luminosity is $400$\,$L_{\odot}$ (purple line).}
\label{ris:radH2O}
\end{figure}

The results of the two-outburst model (Fig.~\ref{ris:radH2O}) demonstrate the complex dynamics of the HDO/H$_2$O ratio. The first powerful outburst ($10\,000 L\odot$) causes significant ice evaporation and changes in isotopic composition, while the subsequent outburst ($400 L_\odot$) further modifies the distribution. At the same time, the HDO/H$_2$O profile in this model is close to that of a single outburst at 10\,000\,$L_{\odot}$ and in agreement with the observational data. The main differences between these models lie in the region inside 50\,au, where the observational data are less reliable. If we compare the two-outburst model with the single-outburst model of 400\,$L_{\odot}$ amplitude, we see the disappearance of the peak at 80\,au, which deviates significantly from the observations. In the fading 2\,000\,$L_{\odot}$  outburst scenario, the HDO/H$_2$O at distances $80-140$\,au is much larger than observed. Similar behaviour is observed in this model at the moment of the luminosity maximum (see Figure~\ref{ris:comb}).

We can conclude that the model with two consecutive outbursts (Fig.~\ref{ris:radH2O}, red line) shows better agreement with the observed HDO/H$_2$O levels, particularly for the stellar luminosity corresponding to $0.1$\,Myr old star. The effect of the past outburst persists in the outer parts of the protoplanetary disc over the considered time interval of $\sim100$\,years. The fraction of deuterated water is determined by the maximum luminosity in the recent past rather than by the current luminosity. In the absence of additional heating mechanisms and assuming an instantaneous change in the thermal structure of the disc, a bright ($\sim10\,000$\,$L_{\odot}$) outburst in the past represents the most appropriate explanation for the observed radial HDO/H$_2$O distribution.

\subsection{Alternative temperature profile}
\label{sec:temperature2}

The key assumption of our modelling is the power-law dependence of the temperature on the radial distance and the luminosity of the central source. We assume the dependence where $T_{\rm mp}\propto L_{\rm acc}^{1/4}R^{-1/2}$, which is derived for fully optically thick protoplanetary discs \citep{2001ApJ...560..957D}. In our choice of the power law exponent $q$ in the radial temperature profile $T_{\rm mp}\propto R^{-q}$ we rely on the observational data on protoplanetary discs that report $q\approx0.5-0.6$ \citep[e.g.][]{2009ApJ...700.1502A}. Similar temperature dependencies are used in many other astrochemical models of protoplanetary discs (e.g., \citet{1999A&A...351..233A}, \citet{2016A&A...595A..83E}, \citet{2025arXiv250617399P}.

However, theoretical studies suggest that a more sophisticated calculation of the thermal structure can be crucial to determine the position of the snowline \citep{2000ApJ...528..995S,2007ApJ...654..606G}.In the models that account for disk flaring, viscous heating, the wavelength dependence of dust opacity, and radiative transfer in the vertical direction for not optically thick discs, the dust temperature in the midplane often follows the relations $T_{\rm mp}\propto L_{\rm acc}^{2/7}R^{-3/7}$ (\citet{1998ApJ...500..411D}, \citet{1999ApJ...527..893D}, \citet{2016A&A...591A..72I}). The weak dependence on luminosity implies that a lower luminosity is sufficient to reach the same sublimation temperature at a given radius.

The use of an alternative temperature dependence could lead to the snowline position to be farther from the star at $\approx400$\,$L_{\odot}$ outburst. We test a model with the temperature defined by the empirical expression from \citet{2016A&A...591A..72I}

\begin{equation}
\label{eq:temp2}
    T_{\rm mp} = 150 \left( \frac{L_{\rm acc} + L_{\star}}{L_{\odot}}\right)^{2/7} \left( \frac{M_{\star}}{M_{\odot}}\right)^{-1/7}\left( \frac{R}{1 {\rm au}}\right)^{-3/7} {\rm K}
\end{equation}

\begin{figure} 
\includegraphics[width=0.8\columnwidth]{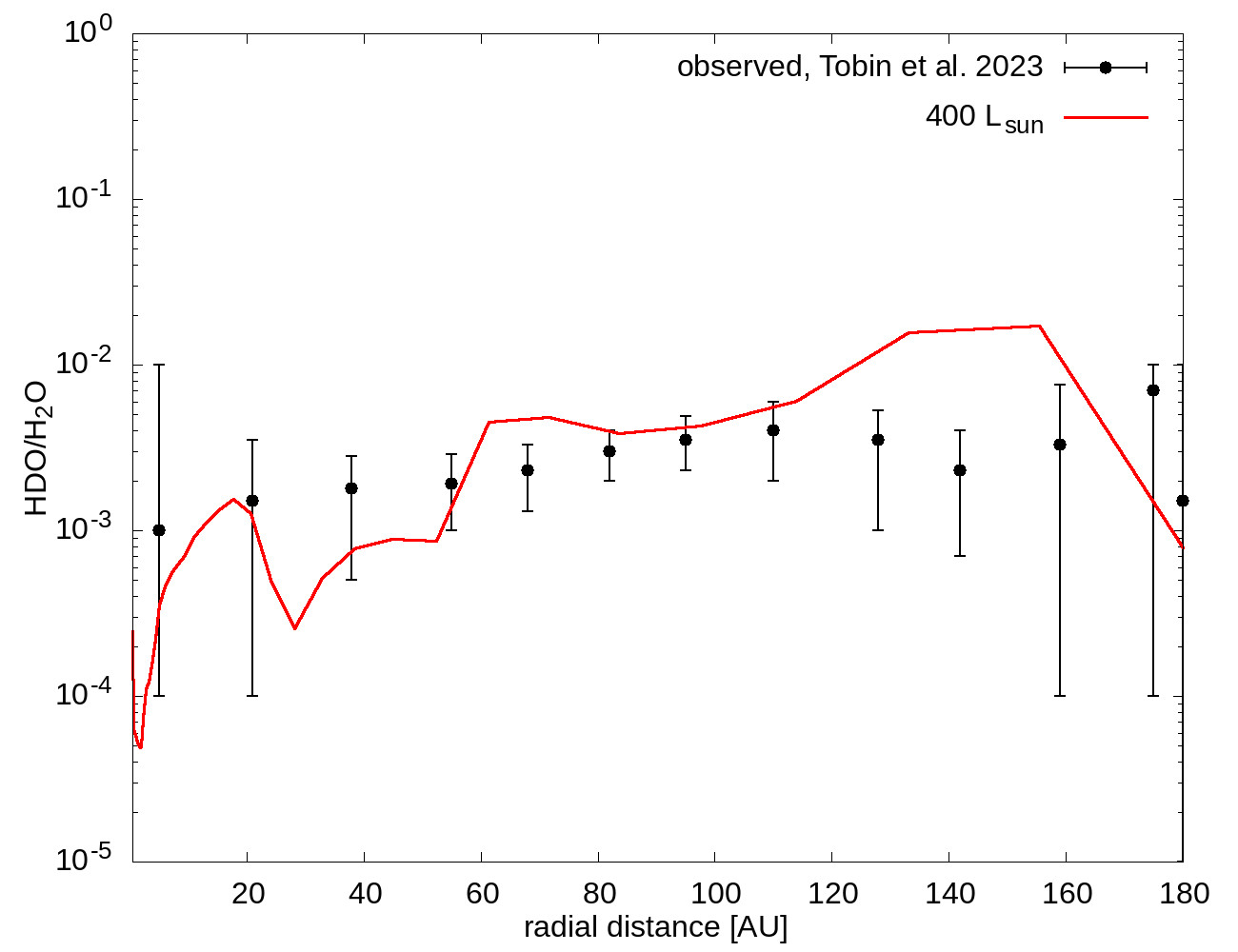}
\caption{Radial profiles of the HDO/H$_2$O ratio in the gas for the model with temperature profile defined by Eq.~\eqref{eq:temp2}. Black markers with error bars show observational data for the V883~Ori protoplanetary disc from \citet{Tobin2023}.}
\label{ris:HDO_H2O_temp2}
\end{figure}

The snowline position in this model is at $\approx50$\,au for the $\approx400$\,$L_{\odot}$ outburst, which is close to the observational limits of $40-100$\,au. The HDO/H$_2$O ratio in this model is shown in Figure~\ref{ris:HDO_H2O_temp2}. It demonstrates a good agreement with the observations, and could be preferred over the results of the models with the classical temperature profile determined by Eq.~\eqref{eq:temp1}. Such temperature structure allows to simultaneously explain the observed snowline position and the profile of HDO/H$_2$O ratio with the current luminosity.

\section{Discussion}
\label{sec:discussion}

The discrepancy between the observed luminosity of 400\,$L_{\odot}$ and the position of the snow lines, as well as the best agreement of the observed HDO/H$_2$O profile with a 10\,000 $L_{\odot}$ model, can be interpreted in different ways.

Modelling indicates that it is the evaporation of water ice as the temperature rises that determines the HDO/H$_2$O ratio in the gas during the outburst in a large part of the disc, for which heating from an outburst of the observed amplitude of 400\,$L_{\odot}$ is insufficient.
This may be due to:
\begin{itemize}
\item Insufficient consideration of additional heating sources (e.g., in the midplane of the disc) or simplified description of the temperature profile.
\item The influence of previous outbursts that changed the initial ice composition.
\item Nonlinear dynamics of chemical processes during outbursts.
\end{itemize}

In order to reproduce the observed high HDO/H$_2$O ratio, it is necessary to transfer these molecules from the ice phase to the gas phase. However, the current accretion luminosity observed in V883~Ori is insufficient to vaporise the water ice in the distance range from~40 to~80~au. The additional temperature rise may arise from a brighter outburst (up to 10\,000\,$L_{\odot}$) that occurred in the recent past and has not been detected. The two-outburst model (10\,000+400 $L_{\odot}$) most closely reproduces the observed HDO/H$_2$O profiles in the V883~Ori disc, allowing us to explain the deuterated water fraction without the need for additional heating sources. The presence of a similar outburst in the past may also help explain the observed position of the water snow line, indirectly determined from the HCO$^+$ emission \citep{2021A&A...646A...3L}. Alternatively, a more sophisticated calculation of the temperature profile \citep[e.g.][]{2016A&A...591A..72I} allows to reproduce the observed snowline position and HDO/H$_2$O profile. Additional uncertainty is added by the time-dependent cooling of the protoplanetary disc after the end of the luminosity outburst: the characteristic heating and cooling times, especially in the dense regions of the disc, can exceed hundreds of years \citep{1997ApJ...490..368C, 2014ARep...58..522V}.

Another restriction on the hypothesis of the bright past outburst is added by the required accretion rate. To reach the accretion luminosity of $10\,000$\,$L_{\odot}$, the mass accretion rate to the star needs to be around $\frac{L_{\rm acc}R_{\star}}{1.5 G M_{\star}}\approx0.8 \cdot 10^{-4}$\,$M_{\odot}$~yr$^{-1}$. During 100\,years of an outburst, around $0.01$\,$M_{\odot}$ will be accreted, which is around 20\% of the adopted disc mass. A shorter duration of the burst could further ease this constraint. However, multiple such outbursts would sufficiently decrease disc mass budget.

The hypothesis that radiative heating may not be the only source of heating in the protoplanetary disc cannot be discarded either. The inclusion of viscous heating may also help to push the water snow line further away from the star, as shown by \citet{2024MNRAS.527.9655A} modelling. At high accretion rates through the protoplanetary disc and in high-density discs, the zone of influence of accretion viscous heating can reach distances of tens of astronomical units \citep{2023A&A...675A.176U}. This mechanism may serve as an additional source of heating, providing an alternative to the bright outburst of luminosity in the past. The expected effect of such heating on the chemical composition is similar to the effect of enhanced radiative heating from an extremely bright luminosity outburst. To distinguish between these scenarios, constraints on the structure of the protoplanetary disc in the vertical direction are needed. Radiative heating primarily affects the upper layers of the disc, while accretion viscous heating is most effective in the disc midplane. Future ALMA observations targeting the highly excited HDO and H$_{2}$O lines, as well as planned mid-infrared ice band spectroscopy using JWST, will test these predictions by constraining the possible vertical distribution of water isotopologues in protoplanetary discs.

In addition to water, numerous deuterated isotopologues of various other molecules, including complex organic molecules, are also observed in the V883~Ori disc. These include deuterated formaldehyde (HDCO), methanol (CH$_2$DOH), acetaldehyde (CH$_3$CDO and CH$_2$DCHO), methyl cyanide (CH$_2$DCN) \citep{2019NatAs...3..314L,2024ApJ...966..119L}. Upper limits on the D/H \citep{2024AJ....167...66Y} ratio were also obtained for some molecules. The diversity of observed complex organics, due to the outburst observed in the gas phase makes this FUor a unique object for studying chemistry in protoplanetary discs. Future more detailed numerical studies of the chemistry of deuterated species in discs around FUors are needed to understand the observed diversity of molecules and the fractionation of deuterium between them.

\section{Conclusions}
\label{sec:conclusion}

In this work, we have numerically modelled the chemical composition of the protoplanetary disc around a young V883~Ori star experiencing a luminosity outburst. We modelled the radial distribution of the deuterated water fraction HDO/H$_2$O and compared it with spatially resolved observational data \citep{Tobin2023} for different luminosity change scenarios and for stellar luminosities corresponding to two ages ($0.1$ and $0.5$ $M_{\odot}$). We show that the closest agreement with observations is achieved in models where the maximum of the luminosity outburst is $\sim10\,000$\,$L_{\odot}$ and $\sim2\,000$\,$L_{\odot}$. At the same time, models with lower luminosities corresponding to the current observed value ($\sim400$\,$L_{\odot}$) show HDO/H$_2$O above the observed value. An agreement of the observed luminosity $\sim400$\,$L_{\odot}$, approximate position of the snowline and the HDO/H$_2$O profile can be reached with the temperature dependence $T\propto R^{-3/7}$ \citep{1998ApJ...500..411D}, which is not seen in the observed temperature profiles of protoplanetary discs \citep{2009ApJ...700.1502A}.

Our calculations show that the interpretation of the observed HDO/H$_2$O distribution profiles is affected not only by the amplitude of the accretion luminosity outburst, but also by the age of the star and its corresponding luminosity. We considered two characteristic ages: $0.1$ and $0.5$\,Myr, corresponding to different stages of protoplanetary disc evolution. The results (see for example, Fig.~\ref{ris:comb}) demonstrate that at higher stellar luminosities (corresponding to a younger age of $0.1$\,Myr) the profile becomes smoother, and the snow line position and water isotopologue abundances are more robust to changes in external conditions. More powerful outbursts such as 2\,000 and 10\,000~$L_{\odot}$, meanwhile, give better agreement with the observed data, especially at ages of $0.1$\,Myr, which may indicate the effect of previous outbursts or a change in the thermochemical structure of the protoplanetary disc with time.

Consideration of the two-outburst model shows that the effect of the brighter first outburst can persist for $\sim100$\,years, leading to an HDO/H$_2$O profile similar to the single-outburst case with $\sim10\,000$\,$L_{\odot}$. This indicates that the system may have experienced a brighter luminosity outburst in the past, as previously proposed by \citet{2021A&A...646A...3L}. This hypothesis is also consistent with the positions of water snow lines determined from observations. However, we cannot exclude that such an effect could be caused by the presence of alternative heating sources instead of the brighter past outburst \citep{2023A&A...675A.176U,2024MNRAS.527.9655A}.

These results emphasise the need to take into account the chemical heterogeneity of ice, the peculiarities of its radial and vertical distribution, and the residual effects of previous chemical evolution when interpreting the observed profiles of water isotopologues in FUor type objects.

\section{Acknowledgments}

The research was carried out under State Assignment FEUZ-2025-0003. 
Tamara Molyarova's work was supported by the Ministry of Science and Higher Education of the Russian Federation, State Assignment No. GZ0110/23-10-IF.

\bibliographystyle{raa.bst}
\bibliography{RAA-2025-0354}

\label{lastpage}

\end{document}